\documentclass[aps, 10pt, prb, twocolumn]{revtex4-1}

\usepackage{hyperref}
\usepackage{gensymb}

\usepackage{graphicx}

\usepackage{color}
\usepackage{setspace}

\begin{document}

\title{Atomic force microscopy reconstruction of complex-shaped\\ chiral plasmonic nanostructures}

\author{A.V. Kondratov, O.Y. Rogov and R.V. Gainutdinov}
\address{
Shubnikov Institute of Crystallography of Federal Scientific Research Centre "Crystallography and Photonics" of Russian Academy of Sciences, 119333 Moscow, Russia}

\email{kondratov@crys.ras.ru}

\vspace{10pt}

\date{September 8, 2016}
%\date{\today}

\begin{abstract}
A significant part of the optical metamaterial phenomena has the plasmonic nature and their investigation requires very accurate knowledge of the fabricated structures shape with a focus on the periodical features. We describe a consistent approach to the shape reconstruction of the plasmonic nanostructures. This includes vertical and tilted spike AFM probes fabrication, AFM imaging and specific post-processing. We studied a complex-shaped chiral metamaterial and conclude that the described post-processing routine extends possibilities of the existing deconvolution algorithms in the case of periodical structures with known rotational symmetry, by providing valuable information about periodical features.
\end{abstract}

\maketitle

\vspace{2pc}

\section{Introduction}

Progress in fabrication of nanostructures and metamaterials \cite{paper_fabrication_2008} allows one to tailor their physical properties different from the properties of their constituent materials \cite{book_metamaterials_2010}. These unique features including enhanced nonlinear response \cite{paper_nonlinear_plasmonics_2012} and optical chirality \cite{paper_chirality_2013} provide opportunities for remarkable applications, e.g. next-generation biosensing \cite{paper_biosensing_plasmonics_2012}. Optical chiral metamaterials are typically built of metallic complex-shaped elements of subwavelength size and achieve up to extreme values of optical chirality \cite{paper_apl_2014}.

Plasmonic nature of a huge part of optical metamaterials makes their investigation even more challenging, since their optical response significantly depends on the nanostructure shape, and a few extra nanometers can noticeably alter observed values \cite{paper_shape_vs_optics_kelly_2003}. Fine nanoscale relief of such structures requires a development of very precise imaging and measurement techniques with high spatial resolution \cite{paper_geometry_distortion_2013}. Hence, there are two suitable methods: atomic-force microscopy (AFM) and scanning electron microscopy (SEM).

An image obtained with SEM is two-dimensional (2D), yet three-dimensional (3D) relief can be recovered using shape-from-shading (SFS) algorithm \cite{paper_shape-from-shading_1999}. It allows one to obtain the shape of the original nanostructure through a non-invasive way \cite{paper_3d_from_sem_2014}. Yet, it's well-known that SEM image depends heavily on both shape and structure material \cite{book_sem_imaging}. Thus, a combination of the SEM and SFS may result in two variants of 3D relief for the same structures fabricated of a different material.

Over the past two decades, AFM became a much more widespread method for the surface relief investigations. The variety of modern AFM techniques, development of microscopes and probe design makes it possible to obtain correct information about a surface relief with a high resolution. Due to a finite size of AFM probe the so-called tip-surface convolution  \cite{paper_afm_theory_1985} should be taken into account when the tip size is equal or exceeds typical sizes of the relief features. There are two certain possibilities to enhance AFM resolution: reducing the tip size and improving post-processing methods.

A considerable number of papers is devoted to the improving of AFM imaging techniques (See e.g. \cite{paper_afm_works_3_1993, paper_canet_2014, paper_afm_works_1_2014, paper_afm_works_2_2015}), especially to the development of algorithms of tip shape deconvolution from AFM images \cite{paper_afm_algos_1_1994, paper_afm_algos_2_1997}. Microscope manufacturers frequently include the suitable routines in an accompanying software. Even the use of deconvolution algorithms is not enough to eliminate the tip distortions completely. Generally, the so-called "blind" surface areas exist at the points where the tip-surface contact is absent, or several contacts occur \cite{paper_keller_1991}. In such a case the deconvolution software is inapplicable, and the combination of special probes and custom data post-processing is necessary.

Ordinary AFM probes with large apex angles are unsuitable for the imaging of the nanostructures with a high aspect ratio due to the blind areas over considerable parts of the structure. Usage of single-wall carbon nanotubes (CNT) as AFM probe proved to be an efficient solution of a tip size problem \cite{paper_carbon_probes_1998}, because their diameter generally does not exceed 1 nm. Though the CNTs help to resolve the fine features of a surface, their usage is generally limited to flat structures, due to the thin probe base and inevitable tip oscillations with multiple contacts inside deep nanostructures with a high aspect ratio. A plausible solution would be to sharpen the probe by the controlled focused ion beam (FIB) etching technique in several steps varying the ion beam current \cite{paper_fib_sharpened_afm_probes_2005}.

In the case of artificial periodic structures of elements with known rotational symmetry the information about a single averaged unit cell shape sometimes is more advantageous than about an entire structure. However, almost all existing AFM techniques and post-processing algorithms are focused on the whole structure shape reconstruction. Here we describe the development of the approach applied to studying of the arrays of chiral holes which have a rather complex shape with certain unknown details \cite{paper_prb_2016}. The main purpose of the AFM shape reconstruction (section \ref{section_methods}) followed by the post-processing routine (section \ref{section_post-pro}) was to obtain a high quality 3D model of the structure unit cell.

\section{Experimental methods}
\label{section_methods}

\subsection{Sample fabrication}
\label{section_sample_fabrication}

\begin{figure}%[h]
\begin{center}
\includegraphics[width=0.48\textwidth]{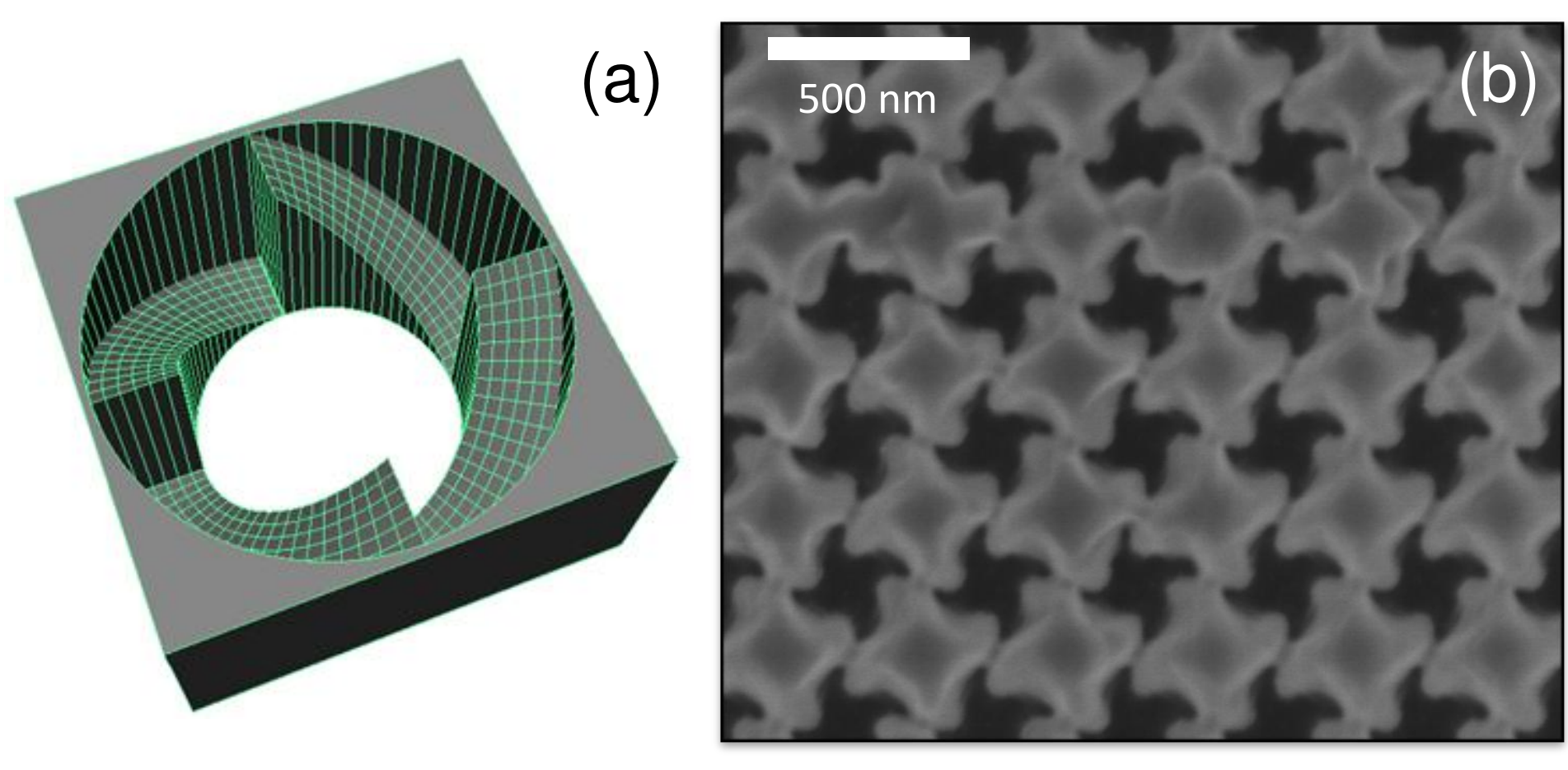}
\caption{FIB digital template (a) \cite{paper_apl_2014} and SEM image (b) of the chiral hole array milled in a 270 nm thick silver foil on a glass substrate.}
\label{figure_sem-image}
\end{center}
\end{figure}

The fabrication of the chiral nanohole arrays was carried out by using a dual-beam FIB system (FEI Scios) controlled via digital templates (figure \ref{figure_sem-image}a). The latter determined the hole shape as a 4-start screw thread circular hole, while the final shape of the fabricated holes is different from those of the template due to the inevitable ion beam defocussing and redeposition of the sample material. A 30 keV gallium beam with current of 50 pA was used as an optimal regime to reduce the influence of these factors. The investigated sample has the period of 360 nm and was milled in a 270 nm thick silver film on a glass substrate (see figure \ref{figure_sem-image}b). The total array size is about 30x30 $\mu m^2$.

\subsection{Spike AFM probes fabrication}
\label{section_methods_probes}

\begin{figure}%[h]
\begin{center}
\includegraphics[width=0.48\textwidth]{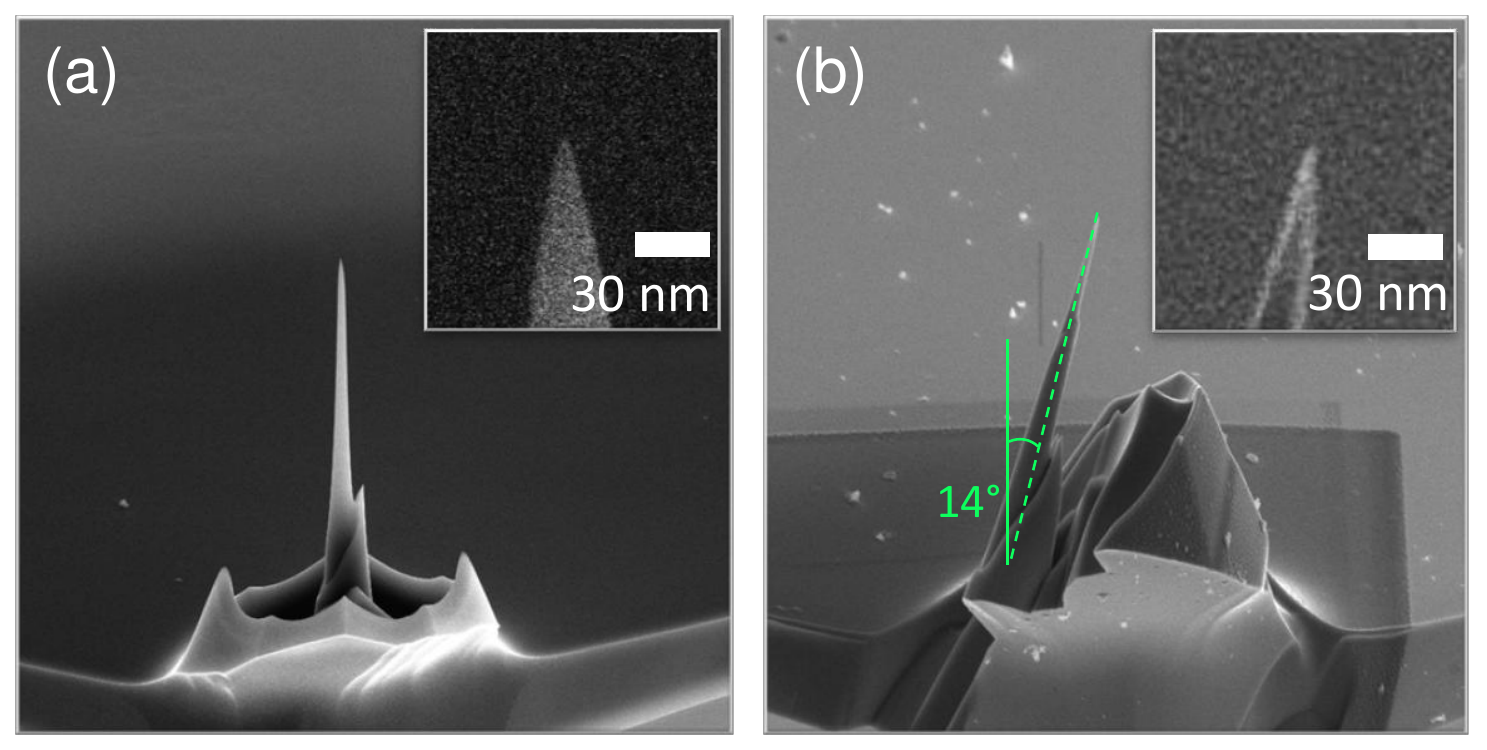}
\caption{SEM images of FIB-sharpened probes: (a) vertical; (b) tilted by $\sim14\degree$, and fragments of the tip (shown as insets).}\label{figure_afm-tips}
\end{center}
\end{figure}

To sharpen ordinary AFM probes the concentric pattern is applied to the tip at the ion beam current value of 3 nA. This operation resulted in a significant change of the aspect ratio from 1:2 to 1:8. We have fabricated custom symmetric vertical (figure \ref{figure_afm-tips}a) and tilted  probes (figure \ref{figure_afm-tips}b) with tilt angle $\sim14\degree$. Silicon probes C21 (Mikroscience, Czech Republic) were chosen as a basis and initially had a tip curvature radius of $R \le 10$ nm, full cone angle $\sim$ $30\degree$, lever stiffness of $k \sim 2.0$ N/m, and resonance frequencies of $f \sim 105$ kHz (B lever).  Having achieved the required probe size, the ion beam current was reduced to 1 pA in several stages, as the pattern dimensions were reduced from 15 $\mu$m to 1 $\mu$m. While the FIB milling is applied to the probe, we leave the inner circular area with the radius of 100 nm intact, in order to achieve the 10 nm tip curvature radius which fits best the chiral nanostructure surface reconstruction task.

\subsection{Imaging techniques}
\label{section_methods_imaging}

All AFM experiments were carried out in a TRACKPORE ROOM-05 control-measuring complex (the range of purity 8 ISO 100). The accuracy of humidity maintaining in the clean room is $\pm1 \%$ in the range $30-70\%$. The accuracy of temperature maintaining is $\pm0.05\ \degree$C in the range $25\pm5 \degree C$. AFM images were obtained in tapping mode with atomic force microscope NT-MDT NTEGRA Prima.

\begin{figure}%[h]
\begin{center}
\includegraphics[width=0.48\textwidth]{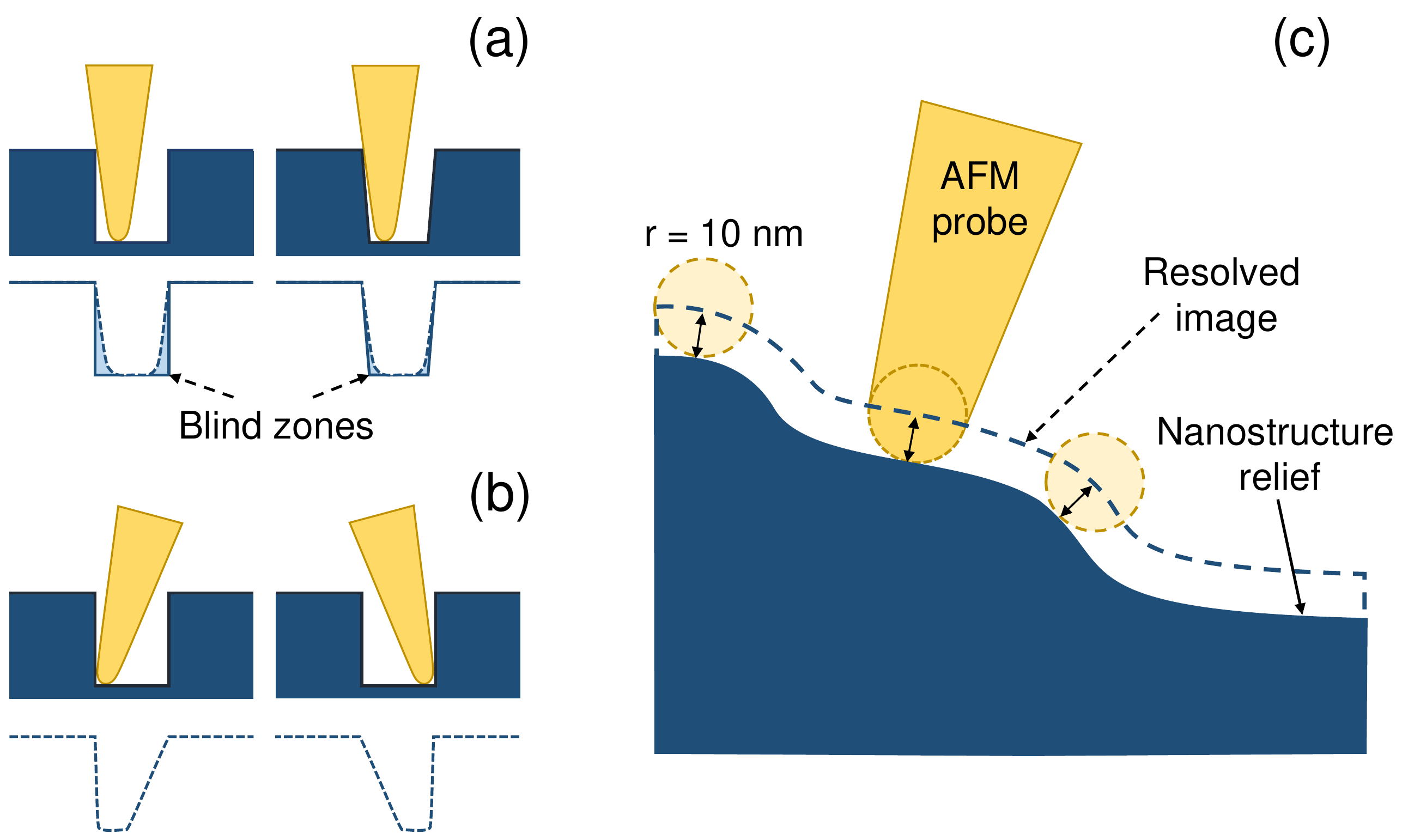}%figure_afm-scheme.png}
\caption{Nanoholes AFM imaging schemes (a-b) and resolved relief shift (c) due to the finite curvature of the tip. Blind zones occur on the holes with a high aspect ratio (a) and can be resolved using tilted probes (b).}
\label{figure_afm-scheme}
\end{center}
\end{figure}

As seen in figure~\ref{figure_afm-scheme}, a vertical AFM probe (figure \ref{figure_afm-scheme}a) produces blind zones during the imaging of deep nanostructures with a high aspect ratio since the multiple contacts occur. To resolve these blind zones the vertical probe can be replaced with a tilted probe that has an appropriate tilt angle. % TODO 
It should be able to resolve the complex topography profile of the one wall of the structure unit element without touching the opposite side (figure \ref{figure_afm-scheme}b).

Unlike the imaging with a vertical probe, the tilted one produces a better result when the surface scanning is performed in the direction opposite to the probe tilt angle. When the scanning course is reversed, the probe behaves like a "plow" and tends to dig under the nanohole edges shifting and making them much deeper than they really are. Thus, a sophisticated topography of the considered chiral nanostructures can be determined by rotating and scanning the sample in direction opposite to the probe tilt, followed by post-processing of the acquired images. This approach benefits from a more detailed image registered due to the ability to resolve the otherwise blind zones.

\section{Data post-processing}
\label{section_post-pro}

\begin{figure}%[h]
\begin{center}
\includegraphics[width=0.48\textwidth]{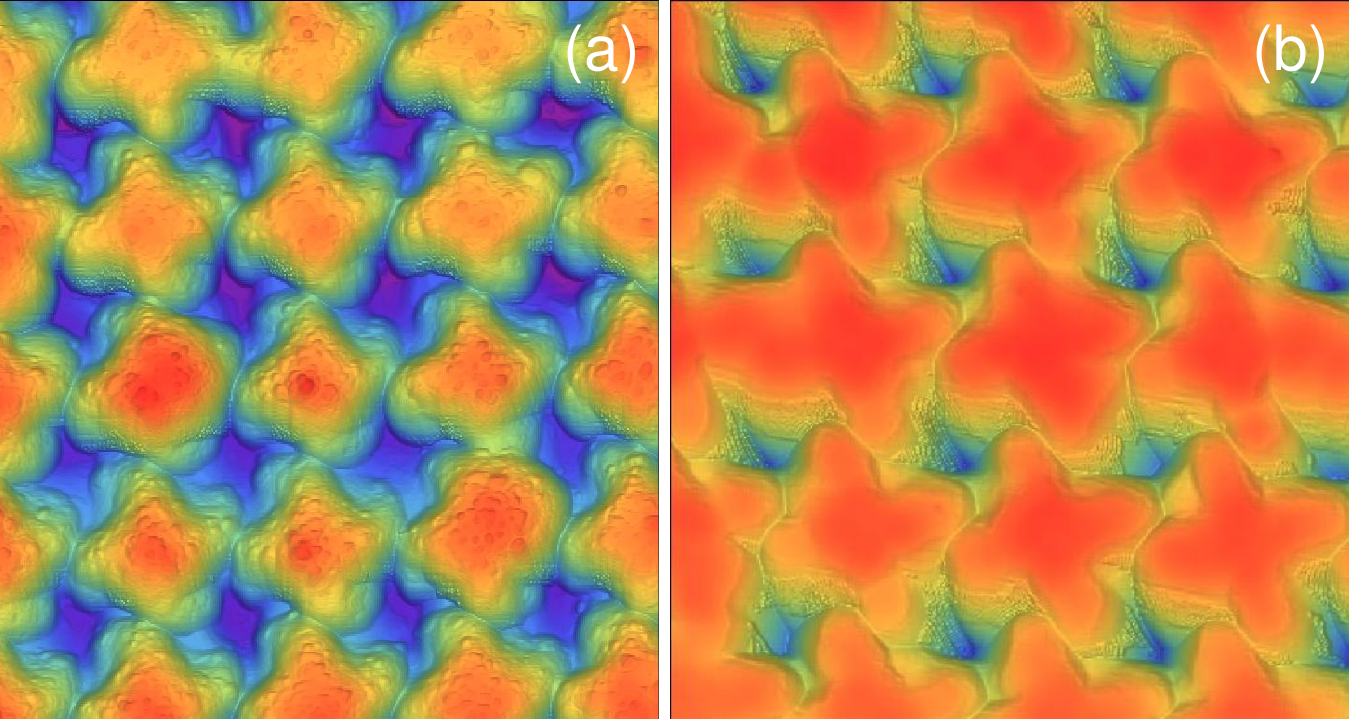}
\caption{Raw AFM data obtained using vertical (a) and tilted (b) probes.}
\label{figure_afm-raw}
\end{center}
\end{figure}

The acquired raw AFM image of the nanohole lattice (figure \ref{figure_afm-raw}) contains a lot of noise, defects and usually is slightly rotated by an arbitrary angle relatively to the image boundaries. These data features complicate further investigation of the structure physical properties, especially theoretical studying of the optical phenomena using a numerical simulation. One cannot perform a full-scale electromagnetic simulation on the entire structure, because it requires enormous computing power, and cannot use a arbitrarily chosen unit cell, because it does not exhibit exact periodicity, rotational symmetry and substantially differs from the remaining cells. Thus, an appropriate post-processing algorithm is required in order to obtain a single averaged unit cell comprising all periodical features of the sample.
 
There are two types of imperfections in the raw AFM data: random noise or fabrication defects and systematic distortions caused by the tip. To extract the periodical features of the structure, and discard random defects and systematic distortions, we developed a numeric post-processing routine which averages over all unit cells in the raw data.  Sharpened probes have a conical form with a tip curvature radius $\sim$10 nm, which systematically makes the holes narrower and more shallow (figure \ref{figure_afm-scheme}c). As an initial post-processing step, the effective tip radius was subtracted from the AFM data along the direction normal to the metal surface. This simple procedure adjusts the hole relief closer to those obtained by SEM imaging. Subsequent part of the algorithm substantially differs for cases of vertical and tilted probes.

\subsection{Vertical probe data}
\label{section_post-pro_vertical}

\begin{figure}%[h]
\begin{center}
\includegraphics[width=0.48\textwidth]{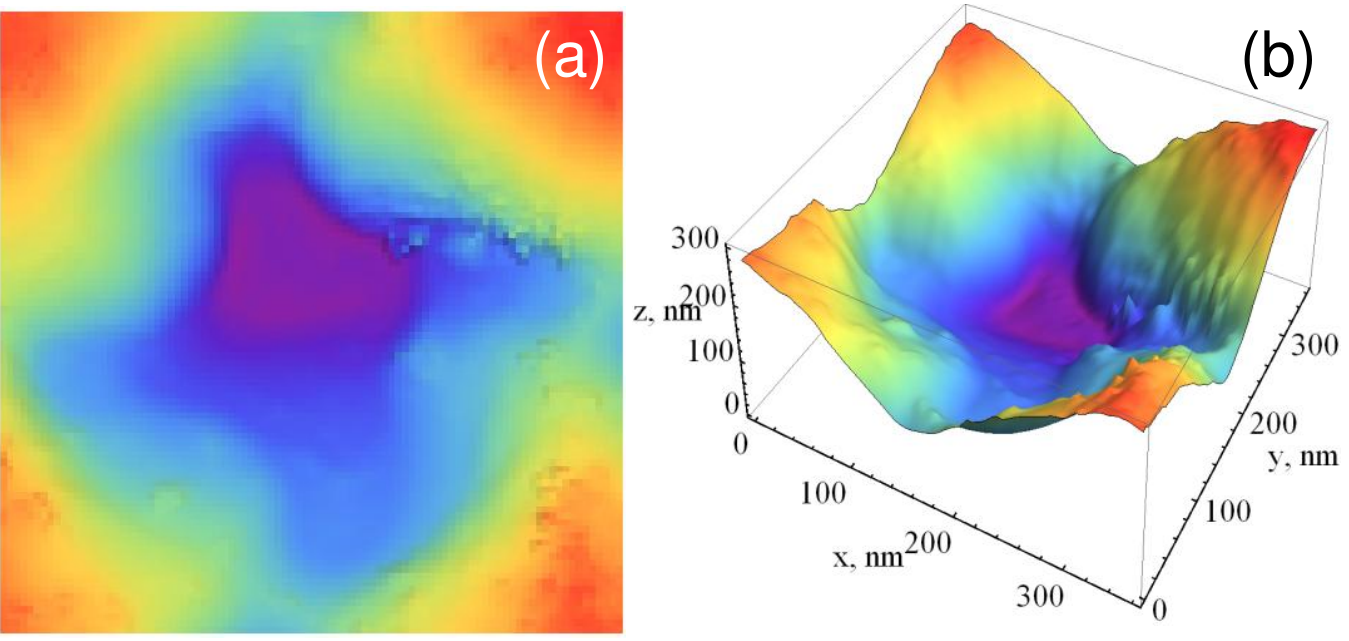}
\caption{Single unit cell from the raw AFM data obtained using vertical probe (a) and corresponding 3D relief (b).}
\label{figure_afm-raw-single-scan-unit}
\end{center}
\end{figure}

\begin{figure}%[h]
\begin{center}
\includegraphics[width=0.48\textwidth]{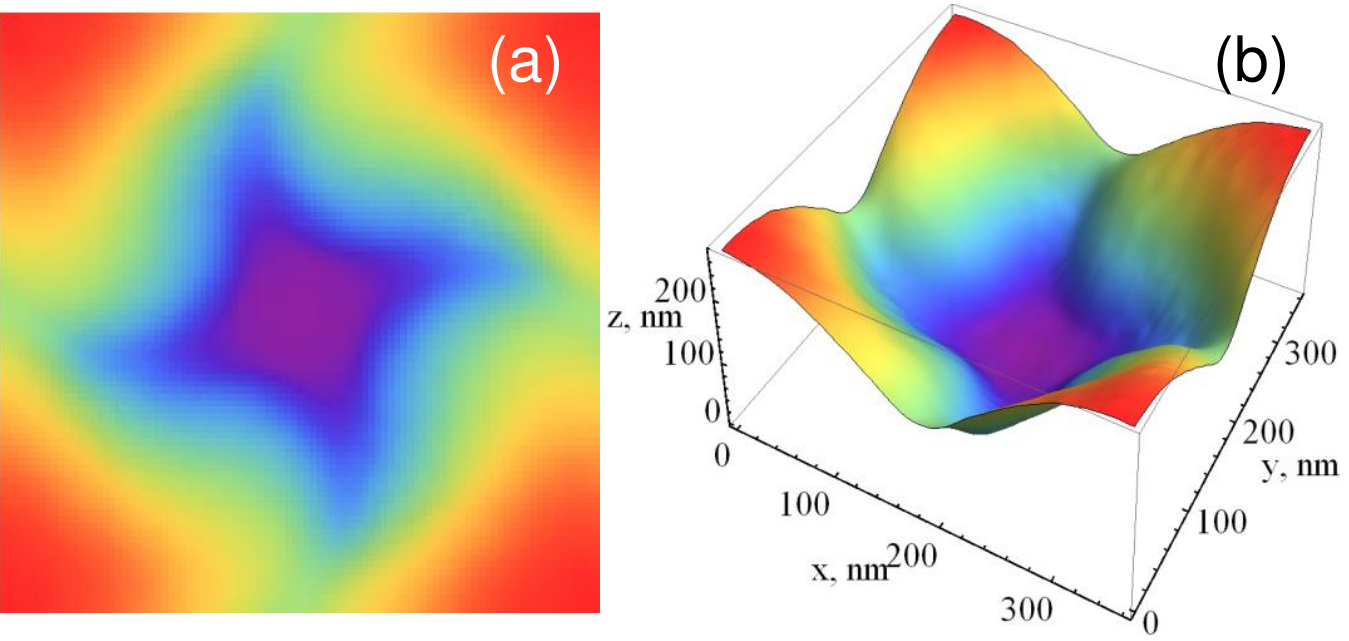}
\caption{Averaged data obtained with the vertical probe (a) and corresponding 3D relief (b).}
\label{figure_afm-average-single-scan-unit}
\end{center}
\end{figure}

Raw AFM data misorientation against image boundaries complicates automatic image processing. Therefore, the acquired structure image should be oriented according to the picture boundaries. The Discrete Fourier Transform was used for the automated detection of the misorientation angle and structure period. It allowed us to rotate the image in the appropriate position and cut entire data into individual unit cell sub-images.

Next, to remove the influence of the random structure defects and AFM noise the averaging was performed in two stages: calculation of the average over all unit cells on the first stage and the same procedure but neglecting the cells with the largest deviations relatively to the average.

Finally, to ensure the precise 4-fold symmetry, the unit cell was averaged with its images rotated by 90, 180 and 270 degrees. Although the resulting mean unit cell (figure \ref{figure_afm-average-single-scan-unit}a) looks smooth and symmetrical, it does not ensure the exact periodicity, because of small mismatches between top-bottom and right-left boundaries. To provide the exact periodicity, the mismatch of the metal surface at the opposite unit cell boundaries was compensated together with its first derivative. The obtained 3D relief of the chiral hole (figure \ref{figure_afm-average-single-scan-unit}b) is strictly periodic, 4-fold symmetric and much smoother than its representative analogue taken directly from the AFM data.

However, even after all steps of the averaging routine the final unit cell hole looks narrower and its walls are more sloping in comparison with the SEM image (compare figure \ref{figure_sem-image}b and \ref{figure_afm-average-single-scan-unit}a).

% TODO: Replace (b) fig with 1D height slice.
\begin{figure}%[h]
\begin{center}
\includegraphics[width=0.48\textwidth]{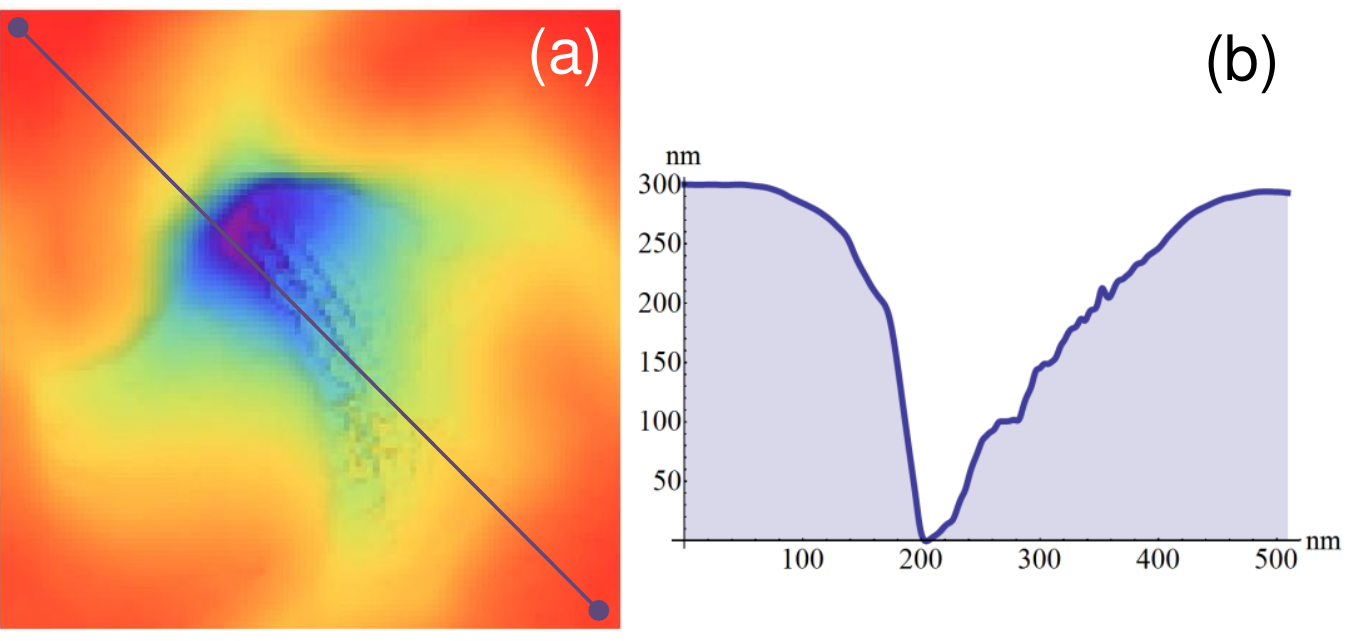}
\caption{Single unit cell from the raw AFM data obtained using tilted probe (a) and sliced profile along the diagonal extraction line (b).}
\label{figure_afm-raw-multi-scan-unit}
\end{center}
\end{figure}

\subsection{Tilted probe data}
\label{section_post-pro_tilted}

Raw AFM data obtained using the tilted probe (see figures \ref{figure_afm-raw}b and \ref{figure_afm-raw-multi-scan-unit}) naturally show an asymmetry in the direction of the probe tilt. It causes several obstacles during post-processing. First, it is harder to automatically find the structure period and detect each unit cell center and boundaries, because the "mass-center" of the structure on the AFM image is shifted relatively to the real geometry center. We overcome it by creating a simple GUI, where one can manually choose the adjacent unit cell centers to find the period, the rotation angle and choose the best cells for a further post-processing.

\begin{figure}%[h]
\begin{center}
\includegraphics[width=0.48\textwidth]{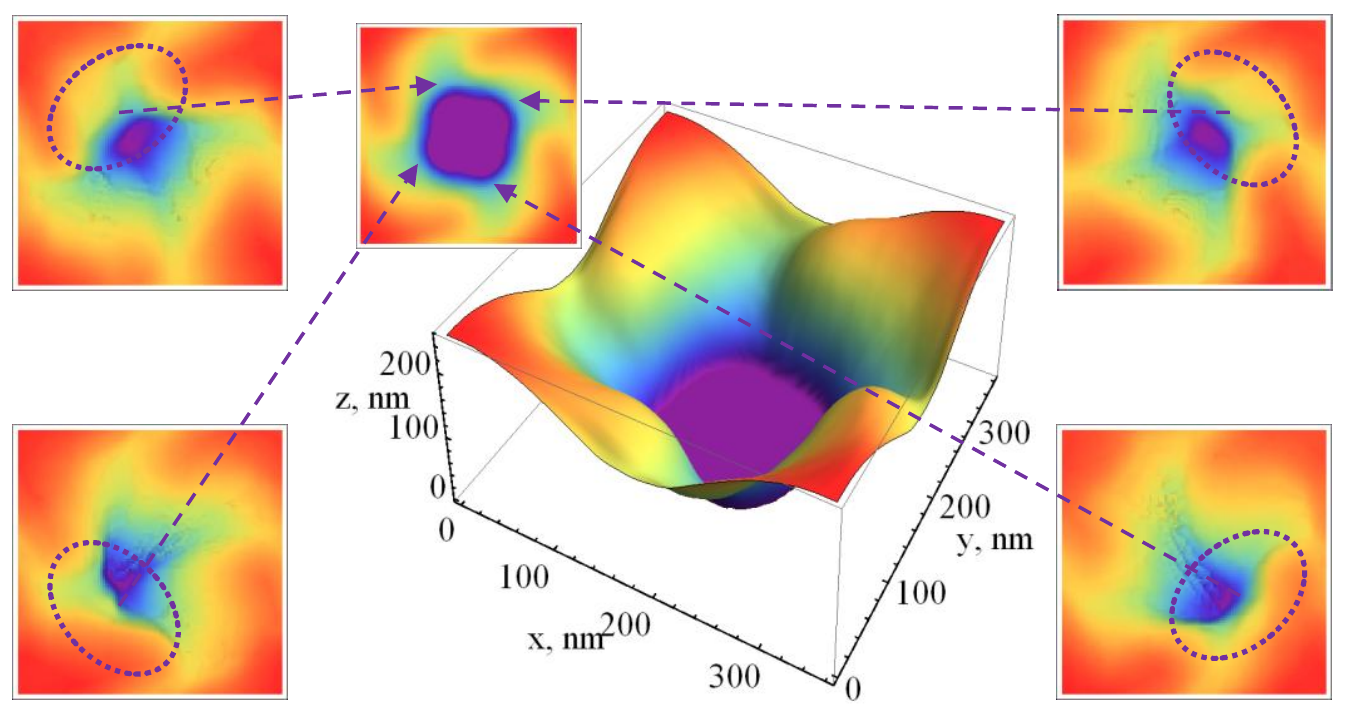}
\caption{Averaging procedure for tilted probe AFM data. Resulting image is combined of finely resolved areas of multiple images.}
\label{figure_afm-average-multi-scan-unit}
\end{center}
\end{figure}

Next, usage of the tilted probe leads to the image where only the wall perpendicular to the AFM probe is resolved with a suitable resolution (figure \ref{figure_afm-raw-multi-scan-unit}). Therefore, for a structure with 4-fold rotation symmetry at least four images are produced. To combine all obtained unit cells we performed weighted averaging, where points with lower height have a greater weight in comparison with higher ones. Therefore, resulting unit cell is patched from four raw cell images (see figure \ref{figure_afm-average-multi-scan-unit}). Remaining 4-fold averaging and boundaries shift was performed in the same as for the vertical probe data in Section \ref{section_post-pro_vertical}. Averaged unit cell has well defined central circular hole with almost vertical walls as seen on the SEM image (compare figures \ref{figure_sem-image} and \ref{figure_afm-average-multi-scan-unit}).

\begin{figure}%[h]
\begin{center}
\includegraphics[width=0.48\textwidth]{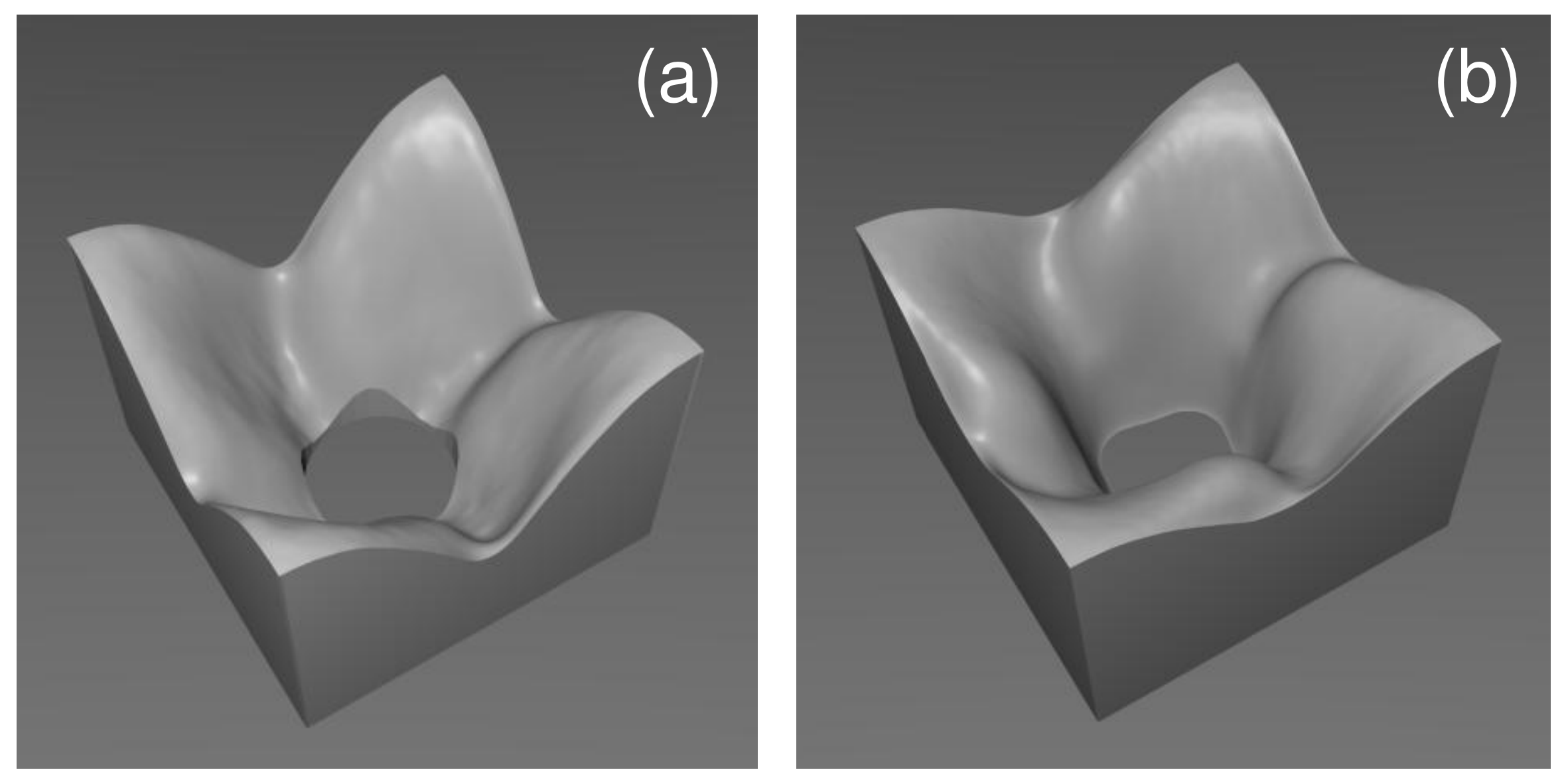}
\caption{Final 3D-models of the unit cell built from the averaged AFM images obtained using the vertical (a) and tilted (b) probes.}
\label{figure_3d-models}
\end{center}
\end{figure}

Resulting unit cell 3D models (see figure \ref{figure_3d-models}) were created using the obtained from the post-processing routine images (see figure \ref{figure_afm-average-single-scan-unit},\ref{figure_afm-average-multi-scan-unit}) as height-maps and open source Blender software. These smooth and strictly periodic unit cell 3D models  can be used in the full-scale FDTD simulation and provide accurate results with a good convergence.

\section{Conclusions}

Usage of the data obtained with the vertical probe leads to a rather sloping relief (figure \ref{figure_3d-models}a). Though we successfully used this model for the experimental results reproduction\cite{paper_prb_2016}, we had to pierce it artificially by a cylinder to make it look more similar to the original structure. The 3D model obtained with the tilted probe combined with the appropriate post-processing routine looks better compared to the one investigated by a vertical probe (figure \ref{figure_3d-models}b). It naturally has almost cylindrical hole and vertical walls.

The described approach shows that the combination of the tilted probe, multi-directional AFM imaging and accurate specific post-processing routine provides an opportunity of the precise shape reconstruction of periodic nanostructures with a high aspect ratio. We developed an implementation of a post-processing algorithm which extends the possibilities of the other deconvolution methods for the purposes of a further theoretical investigation of periodic plasmonic nanostructures, since a single unit cell comprising overall properties is needed instead of an entire structure.

Finally, we note that the tilted probe provides more accurate shape reconstruction results. In some cases, it can resolve well only a part of a complex structure with an appropriate resolution. For example, a periodical nanostructure composed of wide through holes surrounded by a very tiny shallow surface relief (similar to those investigated in the \cite{paper_scientific_reports_2015}). Then a combination of vertical/tilted probes is likely to provide better quality.

\section*{Acknowledgments}

The work was supported by the Russian Science Foundation (Project No. 14-12-00416). We are grateful to IC RAS Shared Research Center (supported by the Ministry of Education and Science of the Russian Federation, Project No. RFMEFI62114X0005) for the equipment provided and to M.V. Gorkunov for valuable advices and criticism.

\section*{References}

\bibliography{afm}

\end{document}